\documentclass{article}
\usepackage{spconf,amsmath,graphicx,hyperref}
\usepackage{titlesec}

\titlespacing*{\subsection}{0pt}{0.4\baselineskip}{0.5\baselineskip}
\usepackage{float}

\usepackage{eso-pic}

\AddToShipoutPictureFG*{%
  \AtPageLowerLeft{%
    \put(0,18){%
      \makebox[\paperwidth][c]{%
        \parbox{0.9\paperwidth}{%
          \centering
          \fontsize{6.5}{7.5}\selectfont
          This work has been submitted to the IEEE for possible publication.
          Copyright may be transferred without notice, after which this version
          may no longer be accessible.
        }%
      }%
    }%
  }%
}

\let\oldthebibliography\thebibliography
\renewcommand{\thebibliography}[1]{%
  \oldthebibliography{#1}%
  \setlength{\itemsep}{1pt}%
  \setlength{\parsep}{0pt}%
  \footnotesize
}

\title{RemixIT-TSE: Progressive Synthetic-to-Real Adaptation for Target Speech Extraction via Target-Aware Supervision and Remixing}

\name{Yu Wang$^{1}$,  Haixin Guan$^{2}$, Shuang Wei$^{1}$, Yanhua Long$^{1*}$ }
\address{$^{1}$Shanghai Normal University, Shanghai, China\\
$^{2}$Unisound AI Technology Co., Ltd., Beijing, China\\
\texttt{wwy97717@gmail.com}  \texttt{, yanhua@shnu.edu.cn}
\thanks{$^{*}$indicates the corresponding author. This work was sponsored by
Natural Science Foundation of Shanghai (Grant No.25ZR1401277).}
\vspace{-0.4cm}}
\begin{document}
%
\maketitle

%

\begin{abstract}

Target Speech Extraction (TSE) in real-world conversational scenarios suffers from severe performance degradation due to the domain gap between synthetic training data and complex acoustic environments, where signal-level ground truth is typically unavailable. To address this challenge, we make the first attempt to extend RemixIT from speech enhancement to TSE and propose a progressive synthetic-to-real adaptation framework for real-world TSE with two fine-tuning stages. The first stage leverages region-wise speaker similarity and silence constraints within a target-aware adaptation framework to jointly optimize the model using synthetic and weakly supervised real-world data, injecting real-world traits while preserving synthetic-learned capabilities. The second stage further adapts the model using only real-world data through our RemixIT-TSE, where quality-filtered teacher pseudo targets, which guarantee reliable student training, provide signal-level supervision via SI-SNR loss. Experiments on the real conversational evaluation set (EVAL-2) of the SLT 2026 REAL-TSE Challenge, the proposed method achieves a 6.53\% relative TER reduction, together with relative improvements of 21.84\% in speaker similarity, 9.89\% in DNSMOS-P808, and 4.10\% in target-activity F1 over the source-domain baseline, demonstrating its effectiveness under unseen real-world conditions. Source code at \url{https://github.com/YuWang-Speech/RemixIT-TSE}.
\end{abstract}
%

\begin{figure*}[!t]
    \centering
    \includegraphics[width=0.98\textwidth]{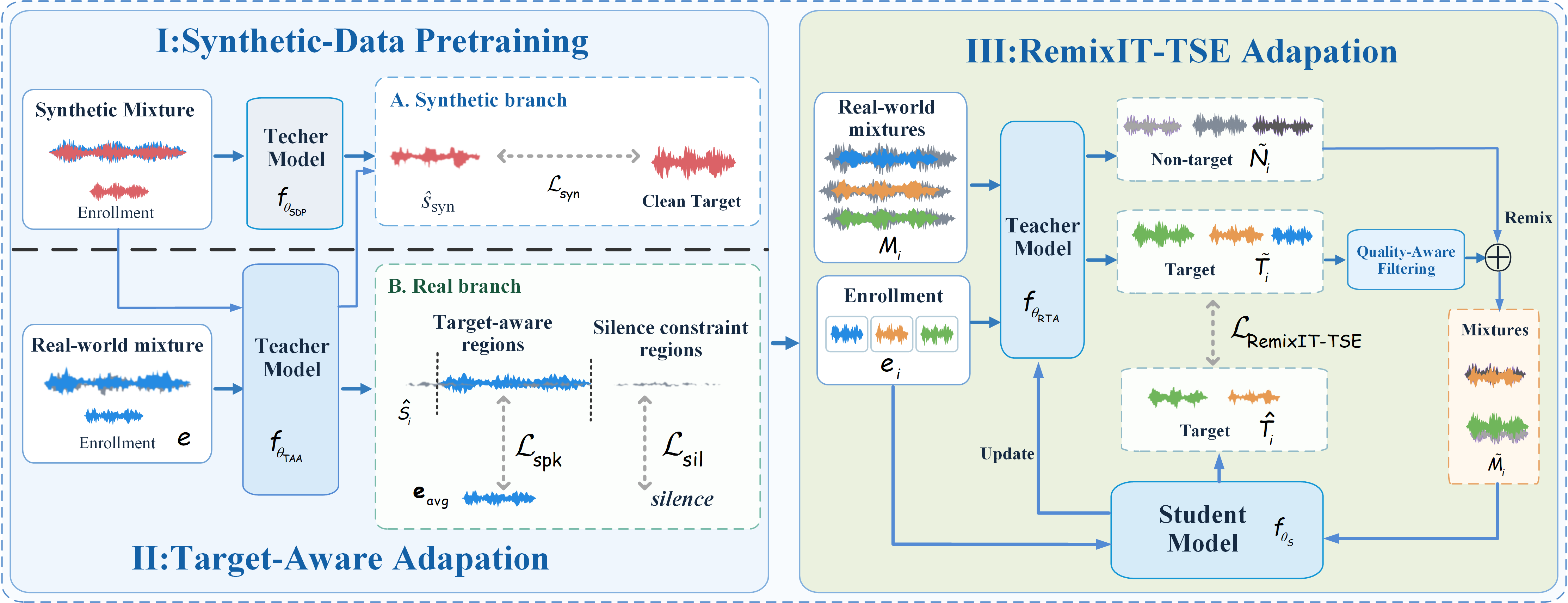}
    \caption{Overview of the proposed RemixIT-TSE framework. Stage I performs supervised synthetic-data pretraining. Stage II jointly adapts the model using synthetic supervision and region-wise speaker and silence constraints on real recordings. Stage III performs real-data-only RemixIT-TSE adaptation using quality-filtered targets, and SI-SNR-based student optimization.}
    \label{fig:framework}
\end{figure*}

\vspace{-1em}
\section{Introduction}
\label{sec:intro}
\vspace{-0.7em}
Target Speech Extraction (TSE) aims to isolate a target speaker's speech from a multi-talker mixture using auxiliary information, typically an enrollment utterance \cite{dpccn,huang2025sef,smma,tselm}. By enabling speaker-specific speech recovery in complex acoustic scenes, TSE is a key front-end technology for real-world applications including multi-party conversations, smart meetings, and speech interaction.

Despite remarkable advances in TSE\cite{dsinet,tang2025lauratse,yang2024target,yangxuetaslp}, most systems are trained on synthetic mixtures such as WHAMR!\cite{wichern2019wham} and Libri2Mix\cite{libri2mix}, where clean target signals are available for supervision. Real conversational recordings, however, contain reverberation, background noise, irregular overlap, and recording-condition mismatch, causing a substantial domain gap and severe performance degradation. Meanwhile, real-world mixtures generally lack signal-level ground truth, but often provide weak annotations such as speaker labels and speaker-attributed, time-aligned transcripts, which can be exploited for real-world adaptation.

Unsupervised and semi-supervised adaptation has been increasingly studied in related speech tasks. MixIT~\cite{mixit} enables unsupervised speech separation from mixtures of mixtures, while RemixIT~\cite{remixit} provides an effective teacher--student paradigm for speech enhancement adaptation without clean target references. However, such adaptation remains largely unexplored for TSE. The closest prior work~\cite{samom} mainly considers noise-free conditions; its extension to noisy scenarios requires additional clean speech, limiting its applicability to real-world recordings where clean target signals are unavailable. 

To address the aforementioned challenges, this paper presents RemixIT-TSE, a novel semi-supervised framework for progressive synthetic-to-real domain adaptation in real-world target speech extraction. The main contributions of this work are summarized as follows:
\vspace{-0.3em}
\begin{itemize}
    \setlength{\itemsep}{0pt}
    \setlength{\parskip}{0pt}
    \setlength{\parsep}{0pt}

    \vspace{-0.3em}
    \item To the best of our knowledge, we are the first to extend RemixIT from speech enhancement to real-world TSE adaptation, bridging the synthetic-to-real domain gap without requiring signal-level clean target references.
    
    \item We design a progressive synthetic-to-real adaptation strategy that first performs mixed-domain weakly supervised fine-tuning with region-wise speaker similarity and silence constraints, and then transitions to real-data-only RemixIT-TSE adaptation.
    
    \item We propose a pseudo-label generation pipeline that constructs enrollments from single-speaker segments and performs quality-aware filtering based on  Speaker Similarity (SIM) and  Token Error Rate (TER).
    
    \item Experiments on the SLT 2026 REAL-TSE Challenge evaluation sets demonstrate consistent improvements in TER, SIM, Mean Opinion Score (MOS), and target-speaker activity F1 over the
source-domain baseline.
\end{itemize}

\vspace{-2.0em}
\section{Proposed Method}
\label{sec:method}

\vspace{-0.9em}
\subsection{Overview of the Framework}
\label{ssec:overview}

The proposed RemixIT-TSE framework progressively adapts a
synthetic-trained TSE model to real-world recordings through three
stages, as illustrated in Fig.~\ref{fig:framework}. Synthetic-Data
Pretraining (SDP) first learns the source-domain model
$f_{\theta_{\mathrm{SDP}}}$ from synthetic data. Target-Aware
Adaptation (TAA) then jointly exploits synthetic supervision and
region-level weak supervision on real recordings to obtain
$f_{\theta_{\mathrm{TAA}}}$. Finally, RemixIT-TSE Adaptation (RTA)
uses only real-world data to obtain $f_{\theta_{\mathrm{RTA}}}$.
Accordingly, training progressively shifts from synthetic-only
supervision to joint synthetic-real adaptation and finally to
real-only adaptation.

\subsection{Enrollment and Pseudo-Target Generation}
\label{ssec:enrollment_pseudo}

Real-world conversational corpora typically lack clean target
references but provide speaker-attributed, time-aligned annotations.
We use non-overlapped single-speaker intervals to construct enrollment
utterances $e$. For target-aware adaptation, five enrollment
utterances from the same target speaker are encoded and averaged to
form the speaker centroid
\begin{equation}
    e_{\rm avg}
    =
    \frac{1}{5}
    \sum_{k=1}^{5}
    \phi(e^{(k)})
\end{equation}
where $\phi()$ denotes the  WeSpeaker ResNet-34 \cite{wespeaker}.

For RemixIT-TSE adaptation, $f_{\theta_{\mathrm{TAA}}}$ initializes
the Stage-III teacher $f_{\theta_{\mathrm{RTA}}}$. Given a real-world
mixture $M_i$ and its enrollment $e_i$, the teacher generates the
target and non-target components as
\begin{equation}
    \tilde{T}_i
    =
    f_{\theta_{\mathrm{RTA}}}(M_i,e_i),
    \
    \tilde{N}_i
    =
    M_i-\tilde{T}_i
\end{equation}

\subsection{Quality-Aware Pseudo-Label Filtering}
\label{ssec:filtering}

Unlike speech enhancement, the pseudo target $\tilde{T}_i$ in TSE may
contain interfering-speaker leakage that is inconsistent with the
target enrollment $e_i$. Such errors may be reinforced during
subsequent remixing. We therefore apply quality-aware filtering to
$\tilde{T}_i$, using SIM as the primary
speaker-identity criterion and TER as a
complementary content criterion:
\begin{equation}
    \mathrm{SIM}_i \geq \tau_{\rm SIM},
    \
    \mathrm{TER}_i \leq \tau_{\rm TER}
\end{equation}
Filtering is applied only to the pseudo target $\tilde{T}_i$; the
corresponding non-target $\tilde{N}_i$ is not independently filtered.
This process filters the real-world recordings to retain a small fraction of high-confidence data for subsequent adaptation.

\subsection{Progressive Synthetic-to-Real Adaptation}
\label{ssec:progressive_adaptation}

Motivated by the insight that a model's foundational extraction capabilities are best established on supervised synthetic data prior to real-world deployment, we propose a progressive synthetic-to-real adaptation framework. The framework consists of the following stages:

\textbf{I: Synthetic-Data Pretraining.}
The model $f_{\theta_{\mathrm{SDP}}}$ is trained using synthetic
mixtures, enrollments, and clean target references. Its supervised objective combines SI-SNR ($\mathcal{L}_{\rm SI\mbox{-}SNR}) $ [14], compressed-magnitude reconstruction($\mathcal{L}_{\rm mag} $), and centroid-based speaker consistency ($\mathcal{L}_{\rm C\mbox{-}SC}$, an advanced variant of target speaker similarity loss) \cite{sc_tse}  :
\begin{equation}
\mathcal{L}_{\rm syn}
=
\lambda_{\rm sisnr}\mathcal{L}_{\rm SI\mbox{-}SNR}
+
\lambda_{\rm mag}\mathcal{L}_{\rm mag}
+
\lambda_{\rm csc}\mathcal{L}_{\rm C\mbox{-}SC}
\end{equation}
where  $\mathcal{L}_\texttt{C-SC}$  loss is defined as:
\begin{equation}
\mathcal{L}_{\rm C\mbox{-}SC}
=
-\log
\frac{
\exp\left(\cos(\phi(\mathbf z),\mathbf c_y)\right)
}{
\sum_{\mathbf c_k\in\mathcal{C}}
\exp\left(\cos(\phi(\mathbf z),\mathbf c_k)\right)
}
\end{equation}
where $\mathbf z$ is the TSE extracted target speech,  $\mathbf c_y$ 
is the enrollment target-speaker centroid , and $\mathcal C$ denotes the set of enrollment target-speaker centroids across the dataset.

\textbf{II: Target-Aware Adaptation.}
In the target-aware adaption stage, the teacher model $f_{\theta_{\mathrm{TAA}}}$ is first initialized by the 
well-trained Stage-I model $f_{\theta_{\mathrm{SDP}}}$, and then further jointly optimized by the synthetic branch $\mathcal{L}_{\rm syn}$ 
and the real branches as shown in Fig.~\ref{fig:framework}. For a real-world mixture, let $\hat{s}_i(t)$
denote the $t$-th frame of the TSE extracted target speech, and let $m_i^{\rm tar}(t)\in\{0,1\}$
denote the target-speaker activity mask derived from the time-aligned
annotations, where $m_i^{\rm tar}(t)=1$ indicates a target-speaker
active region. The corresponding non-target mask is defined as
$m_i^{\rm non\mbox{-}tar}(t)=1-m_i^{\rm tar}(t)$.

Then, the real-branch optimization objective is formulated as the combination of
$\mathcal{L}_{\rm spk}$ and $\mathcal{L}_{\rm sil}$.  $\mathcal{L}_{\rm spk}$  is defined as:
\begin{equation}
\mathcal{L}_{\rm spk} = \mathcal{L}_{\rm C\mbox{-}SC} \left( \phi(m_i^{\rm tar}(t)\hat{s}_i(t)), e_{\rm avg} \right)
\end{equation}
and $\mathcal{L}_{\rm sil}$  is also introduced to enforce silence in the non-target regions as:
\begin{equation}
\mathcal{L}_{\rm sil} = \frac{\sum_t |\hat{s}_i(t)|m_i^{\rm non\mbox{-}tar}(t)}{
\sum_t m_i^{\rm non\mbox{-}tar}(t)}
\end{equation}
Finally, the joint objective loss for the teacher model ($f_{\theta_{\mathrm{TAA}}}$) training is given by:
\begin{equation}
\mathcal{L}_{\rm TAA} =\mathcal{L}_{\rm syn}+\lambda_{\rm r}\left(\lambda_{\rm spk}\mathcal{L}_{\rm spk}+\lambda_{\rm sil}\mathcal{L}_{\rm sil}\right)
\end{equation}

\textbf{III: RemixIT-TSE Adaptation.} 
As illustrated in Fig.~\ref{fig:framework}, Stage-III adopts a teacher-student RemixIT framework~\cite{remixit}. Specifically, the Stage-II model $f_{\theta_{\mathrm{TAA}}}$ serves as the teacher model $f_{\theta_{\mathrm{RTA}}}$ and also initializes the student model $f_{\theta_S}$. Training in this stage  exclusively utilizes real-world data. For real-world mixtures, the teacher model first performs target speaker extraction to 
obtain the estimated target speech recordings $\tilde{T}_i$. 

Compared to the standard RemixIT framework, our approach introduces two key modifications tailored for TSE.
First, we apply a quality-aware pseudo-label filtering mechanism to ensure high-confidence supervision as described in Section \ref{ssec:filtering}. Following this filtering, each selected target--non-target pair 
$(\tilde{T}_i,\tilde{N}_i)$, the non-target components are randomly permuted within the same batch (while strictly avoiding pairing with the same speaker) and remixed to generate bootstrapped mixtures:
\begin{equation}
    \tilde{M}_i = \tilde{T}_i + \tilde{N}_{\pi(i)},
\end{equation}
where $\pi(i)$ denotes the permuted non-target index assigned to the $i$-th target. Then student model $f_{\theta_S}$ takes $\tilde{M}_i$ and the corresponding $e_i$ as input to estimate the target speech:
\begin{equation}
    \hat{T}_i = f_{\theta_S}(\tilde{M}_i, e_i).
\end{equation}

Second, unlike the original RemixIT in \cite{remixit} which aligns both the speech and residual noise 
signals, we focus the optimization solely on the target supervision target speaker consistency. 
Therefore, the student model is optimized against the detached teacher pseudo-target using the negative SI-SNR loss:
\begin{equation}
\mathcal{L}_{\rm RemixIT\mbox{-}TSE} = -\mathrm{SI\mbox{-}SNR}\left(\hat{T}_i,\tilde{T}_i\right).
\end{equation}
Finally, the teacher parameters are updated from the student model via an exponential moving average (EMA) strategy.

\vspace{-1.5em}
\section{Experimental Setup}
\label{sec:experimental}
\vspace{-0.9em}
\subsection{Datasets}
\label{ssec:datasets}

We use WHAMR! and Libri2Mix train-100 as synthetic source-domain datasets for supervised TSE pretraining. For real-world adaptation, we collect conversational recordings from the training portions of AMI~\cite{ami}, AliMeeting~\cite{m2met}, AISHELL-4~\cite{aishell4}, and CHiME-6~\cite{chime6}. We retain mixture segments with both target-speech duration and overlap duration exceeding 5 seconds, resulting in approximately 51.03 hours of real-world mixtures used in the real branch of TAA. These recordings do not provide signal-level clean target references, but contain speaker-attributed, time-aligned annotations. From this corpus, our Quality-Aware Pseudo-Label Filtering extracts 4.2 hours of high-confidence teacher-target data for subsequent RemixIT-TSE adaptation. We evaluate all systems on the development~\cite{realt} and evaluation sets of the 2026 REAL-TSE Challenge~\cite{realtse}. Specifically, \textbf{DEV} and \textbf{EVAL-1} comprise mixture-enrollment pairs from seen source corpora, while \textbf{EVAL-2} consists of pairs from unseen conversational scenarios to assess cross-environment generalization.

\vspace{-0.2em}
\subsection{Configurations}
\label{ssec:implementation}

We adopt USEF-TFGridNet \cite{zeng2025usef}as the TSE backbone, following the original
configuration except for using a causal architecture and 16-kHz
audio. All three stages are optimized using Adam\cite{kingma2015adam}. SDP uses a fixed
learning rate of $1\times10^{-4}$. TAA uses
a learning rate of $1.2\times10^{-5}$ with a 5-epoch linear warmup,
followed by a fixed learning rate. RTA uses a fixed learning rate of
$8\times10^{-6}$ . Gradient clipping \cite{zhang2019gradient}with a
maximum norm of 5 is applied throughout. For TAA, we set
$\lambda_{\rm sisnr}=0.9$,
$\lambda_{\rm mag}=0.05$, and
$\lambda_{\rm csc}=0.1$ in $\mathcal{L}_{\rm syn}$.
For the real-data objective,
$\lambda_{\rm spk}=\lambda_{\rm sil}=0.5$ and
$\lambda_{\rm r}=2.0$.For RTA, we set
$\tau_{\rm SIM}=$  $\tau_{\rm TER}=0.7$. The teacher is updated from
the student using EMA \cite{meanteacher}with momentum of 0.99.

\begin{table*}[!t]
\centering
\caption{Comparison of different adaptation strategies on the REAL-TSE Challenge development set.}
\label{tab:dev_comparison_full}
\scriptsize
\setlength{\tabcolsep}{4pt}
\renewcommand{\arraystretch}{1.3}
\resizebox{\textwidth}{!}{%
\begin{tabular}{clccccccccc}
\hline
ID & Method & TER$\downarrow$ & SIM$\uparrow$ & SIG$\uparrow$ & BAK$\uparrow$ & OVRL$\uparrow$ & P808$\uparrow$ & Precision$\uparrow$ & Recall$\uparrow$ & F1$\uparrow$ \\
\hline
1A & SDP (Baseline) & 0.662 & 0.441 & 2.815 & 2.515 & 2.075 & 2.897 & 0.787 & 0.930 & 0.837 \\
\hline
1B & SDP + SAMoM (synthetic clean) & 0.807 & 0.428 & 1.870 & 2.086 & 1.491 & 2.612 & 0.775 & 0.886 & 0.807 \\
2B & SDP + SAMoM (real single-spk) & 0.753 & {0.499} & 1.811 & 1.668 & 1.421 & 2.726 & 0.762 & \textbf{0.965} & 0.840 \\
\hline
1C & SDP + TAA & 0.661 & 0.459 & \textbf{2.884} & 2.822 & \textbf{2.221} & \textbf{3.103} & 0.772 & 0.949 & 0.848 \\
2C & \textbf{SDP + TAA + RTA (RemixIT-TSE)} & \textbf{0.621} & \textbf{0.501} & 2.544 & \textbf{2.892} & 2.016 & 2.977 & \textbf{0.804} & 0.940 & \textbf{0.854} \\
\hline
\end{tabular}}
\end{table*}

\subsection{Evaluation Metrics}
\label{ssec:metrics}

Following the SLT 2026 REAL-TSE Challenge protocol~\cite{realtse}, we evaluate
TER for target-speech intelligibility, SIM for speaker consistency, DNSMOS P.835 (SIG, BAK, OVRL)~\cite{reddy2022dnsmos} and DNSMOS-P808~\cite{808dnsmos}
for perceptual quality, and target speaker presence rate for temporal
activity prediction. TER is computed using Zipformer ASR~\cite{zipformer} as WER for
English and CER for Mandarin.  Additionally, the target speaker presence rate is measured by temporal Precision, Recall, and F1 using FireRedVAD~\cite{fireredasr2s}.

\vspace{-0.9em}
\section{Results and Analysis}
\label{sec:results}
\vspace{-0.9em}
\subsection{Comparison of Adaptation Strategies}
\label{ssec:adaptation_comparison}

Table~\ref{tab:dev_comparison_full} compares different adaptation strategies on the SLT 2026 REAL-TSE Challenge development set. Starting from the SDP baseline, we evaluate two comparison systems adapted from the noisy-scenario extension of~\cite{samom}: SAMoM (synthetic clean) and SAMoM (real single-spk). Specifically, SAMoM (synthetic clean), which employs clean speech from WHAMR! and LibriMix, degrades TER from 0.662 to 0.807 and P808 from 2.897 to 2.612. Replacing it with SAMoM (real single-spk), which utilizes 5.1 hours of filtered real single-speaker speech, improves SIM to 0.505, but TER and P808 still degrade to 0.753 and 2.726. This indicates that directly extending remix-based semi-supervision to real recordings is difficult, since real mixtures are substantially more complex and the available single-speaker segments are not signal-level clean target references.

In contrast, TAA preserves the baseline TER while improving P808 from 2.897 to 3.103 and F1 from 0.837 to 0.848, providing a stable synthetic-to-real transition. Subsequent RTA further reduces TER to 0.621 and improves SIM and F1 to 0.501 and 0.854, respectively, showing the benefit of progressive adaptation.

\begingroup
\setlength{\intextsep}{3pt}
\setlength{\textfloatsep}{3pt}
\setlength{\floatsep}{3pt}

\vspace{-0.5em}
\begin{table}[H]
\centering
\caption{Performance comparison on the REAL-TSE Challenge evaluation sets.}
\label{tab:main}
\footnotesize
\renewcommand{\arraystretch}{1.2}
\setlength{\tabcolsep}{2.5pt}
\resizebox{\columnwidth}{!}{%
\begin{tabular}{llccccc}
\hline
Set & System & TER$\downarrow$ & SIM$\uparrow$ & OVRL$\uparrow$ & P808$\uparrow$ & F1$\uparrow$ \\
\hline
EVAL-1 & SDP  & 0.726 & 0.485 & 2.049 & 2.923 & 0.824 \\
EVAL-1 & \textbf{RemixIT-TSE} & \textbf{0.680} & \textbf{0.533} & \textbf{2.173} & \textbf{3.088} & \textbf{0.837} \\
\hline
EVAL-2 & SDP  & 0.763 & 0.335 & 1.850 & 2.710 & 0.804 \\
EVAL-2 & \textbf{RemixIT-TSE} & \textbf{0.713} & \textbf{0.408} & \textbf{2.040} & \textbf{2.978} & \textbf{0.837} \\
\hline
Overall & SDP  & 0.748 & 0.395 & 1.929 & 2.790 & 0.812 \\
Overall & \textbf{RemixIT-TSE} & \textbf{0.700} & \textbf{0.458} & \textbf{2.093} & \textbf{3.022} & \textbf{0.837} \\
\hline
\end{tabular}}
\end{table}
\endgroup

\vspace{-0.0em}
\subsection{Results on REAL-TSE  Evaluation Sets}
\label{ssec:main_results}
\vspace{-0.2em}
Table~\ref{tab:main} reports the performance comparison on the REAL-TSE Challenge evaluation sets. RemixIT-TSE consistently outperforms the SDP baseline across both seen (EVAL-1) and unseen (EVAL-2) conditions. Specifically, on the EVAL-1, our method yields solid improvements across all perceptual and tracking metrics. More importantly, on the challenging unseen EVAL-2, RemixIT-TSE achieves substantial gains, such as reducing TER from 0.763 to 0.713 and boosting P808 by 0.268, demonstrating its robust cross-environment generalization capability. Overall, across the combined evaluation set, the proposed approach demonstrates comprehensive performance improvements, validating the effectiveness of our progressive adaptation framework.

\vspace{-1.5em}
\begin{table}[H]
\centering
\caption{Ablation study of the modifications to RemixIT for RemixIT-TSE on the REAL-TSE development set.}
\label{tab:filter_ablation}
\small
\setlength{\tabcolsep}{0.1pt}
\renewcommand{\arraystretch}{1.3}
\begin{tabular*}{\columnwidth}{@{\extracolsep{\fill}}lcccccc@{}}
\hline
Configuration & Data (hrs) & TER$\downarrow$ & SIM$\uparrow$ & OVRL$\uparrow$ & P808$\uparrow$ & F1$\uparrow$ \\
\hline

\textbf{RemixIT-TSE} & \textbf{4.2} & \textbf{0.621} & \textbf{0.501} & \textbf{2.016} & \textbf{2.977} & \textbf{0.854} \\
\hline
w/ residual loss & \textbf{4.2} & 0.752 & 0.498 & 1.421 & 2.714 & 0.841 \\

w/o filtering & 51.03 & 0.703 & 0.449 & 1.450 & 2.649 & 0.792 \\

\hline
\end{tabular*}
\end{table}

\vspace{-0.9em}
\subsection{Analysis of Modifications for \mbox{RemixIT-TSE}}
\label{ssec:remixit_analysis}

Table~\ref{tab:filter_ablation} verifies the two key modifications
for adapting RemixIT to TSE. Without filtering, using 51.03 hours of
real data degrades all metrics, whereas only 4.2 hours of filtered high-confidence data, with merely 471 iterations, achieves the best performance, showing that
pseudo-target quality is more important than data quantity. Moreover,
restoring the original residual supervision also degrades performance,
as the non-target component may contain interfering speakers, noise,
and teacher errors. Therefore, RemixIT-TSE uses quality-filtered
teacher targets with target-only supervision.

\vspace{-1.5em}
\section{Conclusion}
\label{sec:conclusion}
\vspace{-0.8em}
This work introduces RemixIT-TSE, 
 first extend the RemixIT paradigm from speech
enhancement to real-world target speech extraction, with two key
modifications: quality-aware pseudo-target filtering and target-only
supervision. By combining weak
speaker-attributed annotations with target-aware adaptation and
quality-filtered pseudo-target learning, the proposed progressive
synthetic-to-real adaptation gradually transfers a synthetic-trained
TSE model to real recordings without requiring signal-level clean
target references. Experiments on the REAL-TSE Challenge demonstrate
consistent improvements in extraction accuracy, speaker similarity,
perceptual quality, and target-speaker activity prediction. Future
work will investigate extensions to more challenging conditions,
including extremely noisy environments, 
and scenarios with dynamically changing target and interfering
speakers.

\bibliographystyle{IEEEtran}
\bibliography{strings,refs}

\end{document}